\documentclass{aa}
\usepackage{graphicx}
\usepackage{txfonts}
\usepackage{amsmath}
\usepackage{booktabs}
\usepackage{hyperref}
\begin{document}

   \title{TDCOSMO XIV: Practical Techniques for Estimating External Convergence of Strong Gravitational Lens Systems and Applications to the SDSS J0924+0219 System}

   \author{Patrick Wells
          \inst{1}
          \and
          Christopher D. Fassnacht\inst{1}
          \and
          C. E. Rusu\inst{2}
          }

   \institute{Department of Physics and Astronomy, University of California, Davis, CA 95616, USA\newline
   email: \texttt{pwells@ucdavis.edu}
   \and
   National Astronomical Observatory of Japan, Tokyo, 181-8588, Japan}
   \date{Received Feb. 6th, 2023; accepted -- --, ----}

 
  \abstract
   {Time-delay cosmography uses strong gravitational lensing of a time-variable source to infer the Hubble Constant. The measurement is independent from both traditional distance ladder and CMB measurements. An accurate measurement with this technique requires the consideration of the effects of objects along the line of sight outside the primary lens, which is quantified by the external convergence ($\kappa_{\rm{ext}}$). In absence of such corrections, $H_0$ will be biased towards higher values in overdense fields and lower values in underdense fields.}
   {We discuss the current state of the methods used to account for environment effects. We present a new software package built for this kind of analysis and others that can leverage large astronomical survey datasets. We apply these techniques to the SDSS J0924+0219 strong lens field.}
   {We infer the relative density of the SDSS J0924+0219 field by computing weighted number counts for all galaxies in the field, and comparing to weighted number counts computed for a large number of fields in a reference survey. We then compute weighted number counts in the Millennium Simulation and compare these results to infer the external convergence of the lens field.}
   {Our results show the SDSS J0924+0219 field is a fairly typical line of sight, with median $\kappa_{\rm{ext}} = -0.012$ and standard deviation $\sigma_{\kappa} = 0.028$.} 
   {}


   \maketitle
%

\section{Introduction}

One of the most important problems in modern cosmology is the so-called Hubble Tension: the name given to an apparent discrepancy between the value of the Hubble Constant ($H_0$) inferred from the Cosmic Microwave Background assuming the standard $\Lambda$-CDM cosmological model \citep[e.g.][]{planck_result}, and the value inferred from Cepheid-calibrated Type Ia supernovae. \citep[e.g.][]{Riess_2021_hubble}. The solution to this discrepancy may involve unknown systematics, new physics, or some combination of the two. Work on the resolution is ongoing, but having independent methods for inferring the value of the Hubble constant is critical for solving the problem. Such methods include using the tip of the red giant branch to calibrate supernovae,\citep{trgb_h0} and BAO+BBN \citep{bao_h0}, among others. Here we consider Time Delay Cosmography, which uses strong gravitational lensing of time-variable sources (usually quasars) to infer the value of $H_0$. 

TDCOSMO is an international collaboration which aims to use strong gravitational lensing to infer the value of $H_0$ with sub-percent precision. A typical lens in the TDCOSMO sample involves a source quasar being strongly lensed by a foreground galaxy, producing four images. When the quasar's luminosity varies, this variation appears in each of th eimages at different times. The "time delay" between two images depends on the mass distribution of the lens, and the \textit{time-delay distance}:

\begin{equation} \label{ddt}
D_{\Delta t} = (1 + z_{\rm{d}})\frac{D_{\rm{d}} D_{\rm{s}}}{D_{\rm{ds}}} \propto \frac{1}{H_0}
\end{equation}

where $z_{\rm{d}}$ is the redshift of the lensing galaxy (or "deflector"), and $D_{\rm{d}}, D_{\rm{s}}$, and $D_{\rm{ds}}$ are the angular diameter distances to the deflector, source, and from the deflector to the source, respectively. Given a model of the lens and a measurement of the time delays, it is possible to infer the time-delay distance and therefore $H_0$.

When analysing a strong lens, there are a number of sources of uncertainty that propagate to the final inferred value of $H_0$ \citep[see][for a more complete overview]{tdcosmo1}. Among these are so-called environmental effects: gravitational bodies besides the primary lens that affect the lensing observables and, by proxy, the inferred value of $H_0$. Accounting for these effects is crucial in improving the precision of the inferred measurement. Broadly speaking we treat weak perturbers differently from strong ones, with the difference being determined with the "flexion shift" formalism \citep[see for example ][]{strides_env, Sluse_2019a, flexion_formalism}.

In this paper, we focus on the techniques for inferring the cumulative effect of all weak perturbers along the line of sight to the source quasar. In principle, an "ideal" analysis would include these perturbers in a complete mass model of the system, rendering the analysis we do here unnecessary. While this may be possible for a few systems where high-quality spectroscopic data of the perturbers is available, it does not scale well to the large number of strong lensing systems we aim to analyze in the future. Instead of building a complete model, we compare the field of interest to some suitably large reference field, with the goal of estimating the relative density of the field as compared to the universe at large. This is a purely statistical analysis. In this paper, we will discuss the current state of this technique, apply it to the field around the SDSS J0924+0219 lens system (hereafter J0924, see section \ref{lens_field_sec}), and discuss how it might be iterated upon in the future. The cumulative effect of all weak perturbers is parameterized by the \textit{external convergence}, denoted $\kappa_{\rm{ext}}$.

 While controlling uncertainties on individual lens systems is crucial, the precision of the $H_0$ measurement can be improved by including many lenses in the final inference. Of course, this is much easier said than done as the amount of work required to fully analyze a single lens is significant. However in the modern era of data-driven astronomy, the amount of available data is increasing by orders of magnitude and developing tools for efficiently analyzing these datasets is a top priority. To that end, we introduce \texttt{lenskappa}, a new package designed specifically for the types of environment analysis discussed in this paper, but with broader applications. \texttt{lenskappa} is built on top of \texttt{heinlein}, a data management library designed for use with large astronomical survey datasets. At present \texttt{heinlein} and \texttt{lenskappa} support Subaru Hyper Suprime Cam Strategic Survey Program \cite[HSC-SSP;][]{hsc_dr2}, the Dark Energy Survey \citep{des_design}, and the CFHT Legacy Survey \citep{cfhtls_results}. 

This paper is organized as follows: In Section 2, we outline the fundamentals of Time Delay Cosmography and the basic process used to compute a value of $\kappa_{\rm{ext}}$ for a generic lens system. In Section 3 we discuss the datasets used in our analysis, and introduce \texttt{lenskappa}. In Section 4, we report the results of our analysis of the J0924 field obtained using \texttt{lenskappa}. In Section 5 we discuss our results and look forward to later work.

\section{Summary of the technique}\label{overview}

 At a high level, gravitational lensing is the result of the underlying mass distributions of the universe, with greater concentrations of mass resulting in more significant lensing. In general, it is difficult to assess the impact of any one mass structure on the image of some background object. Strong lenses are naturally an exception, which occurs when a single, high-mass object falls on (or nearly on) the axis drawn between an observer and some background source. However we are seeking to understand the total lensing effect of many additional objects, each producing a tiny effect on the lensing observables. As light passes through the universe, the amount of lensing it undergoes will be determined by the relative concentration of mass along its full path of travel. Our analysis therefore seeks to infer the effects of the mass distributions in our line of sight by comparing it to many lines of sight in the universe. By determining how close this line of sight is to the average line of sight in the universe, we can place constraints on its impact on our primary lens observables. In this section, we introduce the basics of time delay cosmography and describe the technique we use to estimate $\kappa_{\rm{ext}}$ for a generic lens system. For a more complete overview, we refer the reader to \citet{tdc_overview} and \citet{birrer_summary}.

\subsection{Fundamentals of Time Delay Cosmography, and Strong vs. Weak Lensing}

Time delay cosmography focuses on strong lensing of time-variable sources, usually a quasar. As the luminosity of the source varies, this variation appears in each of the several images of the source, but not at the same time. The time delay between any two images can be written as follows:

\begin{equation} \label{deltat}
\Delta t_{ab} = \frac{D_{\Delta t}} {c}\bigg[ \frac{(\overrightarrow{\theta_a} - \overrightarrow{\beta})^2}{2} - \frac{(\overrightarrow{\theta_b} - \overrightarrow{\beta})^2}{2} - \psi(\overrightarrow{\theta_a} ) + \psi(\overrightarrow{\theta_b})\bigg],
\end{equation}

where $\overrightarrow{\theta}$ represents the angular position of an image on the sky, $\overrightarrow{\beta}$ represents the actual (unobservable) angular position of the source on the sky, and $\psi$ is the scaled lensing potential. The first two terms are a result of the different distances traveled by the light for the two images, while the later two terms are the difference in the Shapiro time delay. The goal of a complete cosmographic analysis is to infer the \emph{time-delay distance} (see eq. \ref{ddt}). Using measurements of the time delay combined with a robust model of the lensing galaxy it is possible to measure the time delay distance and, therefore, the Hubble constant.

The multiple images observed in such a system are en example of \textit{strong gravitational lensing}. Qualitatively, strong lensing is any lensing which produces multiple images of some background object. Quantitatively, strong lensing occurs whenever the local density of the lens is greater than the \textit{lensing critical density:}

\begin{equation}
    \Sigma_{cr} = \frac{c^2 D_s}{4\pi GD_{ds}{D_d}}
\end{equation}

For a given perturber, the \textit{convergence} at a given location is defined as the local density in units of the critical density:

\begin{equation}
    \kappa(\overrightarrow{\theta}) = \frac{\Sigma(\overrightarrow{\theta})}{\Sigma_{cr}}
\end{equation}

For $\kappa < 1$, strong lensing does not occur. Instead, mass distributions with $\kappa < 1$ result in magnification or demagnification of the image of the background object. This is the situation for the perturbers along the line of sight in our system that are not included in the primary lens model, though we note that some perturbers that meet this criterion \textit{are} included in the mass model based on their flexion shift (see section \ref{flexion_section}), and therefore excluded from our statistical analysis. However, the effect of the pertubers we \textit{do} include in the estimate discussed here is not directly observable because the actual angular size of the background object is not known.

We quantify the cumulative effect of all perturbers (not including those incorporated into the primary lens model) and voids along the line of sight with the external convergence, $\kappa_{\rm{ext}}$. Conceptually, the value of $\kappa_{\rm{ext}}$ is the convergence of a mass sheet which, if placed coplanar to the primary lens, would produce the same magnification or demagnification as the perturbers do collectively. The external convergence is defined relative to that of a line of sight where the mass distribution is smoothly distributed with a density equal to the global mass density of the Universe.  Therefore positive values of $\kappa_{\rm{ext}}$ represent lines of sight that are overdense with respect to the overall density of the Universe,  while underdense lines of sight have negative values.

These perturbers can also produce shear, denoted $\gamma_{\rm{ext}}$, which results in stretching and distorting of the image of the source. However, this effect can be estimated in the primary lens model, as it affects the location of the the images and the shape of the Einstein ring (if present). We use the $\gamma_{\rm{ext}}$ constraint from the lens model in our analysis as an additional constraint (see section \ref{inference})

Assuming $\kappa_{\rm{ext}}$ can be measured, it serves as a correction factor to the computed value of the time delay distance:
\begin{equation} \label{ddt_correction}
D_{\Delta t} = \frac{D'_{\Delta t}}{1-\kappa_{\rm{ext}}},
\end{equation}
where $D'_{\Delta t}$ represents the uncorrected value of the time delay distance. This propagates directly to the inferred value of the Hubble constant by

\begin{equation} \label{hubble_correction}
H_{0} = (1 - \kappa_{\rm{ext}}) H'_0
\end{equation}

\noindent where $H_0'$ is the value of the Hubble Constant inferred before correcting for the environment. We see therefore that in the absence of the appropriate corrections, the inferred value of Hubble constant would be biased towards higher values in an overdense field, and lower values in an underdense field.

We now review the techniques we use to compute the value of $\kappa_{\rm{ext}}$ for a generic lens system.

\subsection{Relevant Perturbers in the Line of Sight}\label{flexion_section}

To start, we identify objects along the line of sight that contribute to $\kappa_{\rm{ext}}$. We separate objects into \textit{strong} and \textit{weak} perturbers, using an operational definition discussed below. Strong perturbers are generally close to the center of the field or agalaxy group \citep[see for example][]{Sluse_2019a, Fassnacht_2006_env}. Strong perturbers are included explicitly in the mass model of the lens, while weak perturbers are treated statistically. To separate these, we use the \textit{flexion shift} formalism, first proposed in \cite{flexion_formalism}. The flexion shift of an object is given by

\begin{equation}
    \label{flexion_eqn}
    \Delta_3 x = f(\beta)\times \frac{(\theta_E \theta_{E,p})^2}{\theta^3}
\end{equation}

where $\theta_E$ and $\theta_{E,p}$ are the Einstein radii of the main lens and perturber respectively, and $\theta$ is their angular separation. The quantity $f(\beta)$ is given by

\begin{equation}
f(\beta) = 
\left\{
    \begin{array}{lr}
        (1-\beta)^2, & \text{if } z > z_{\rm{d}}\\
        1, & \text{if } z < z_{\rm{d}}
    \end{array}
\right\}
\end{equation}
where

\begin{equation}
    \beta = \frac{D_{\rm{dp}}D_{\rm{s}}}{D_{\rm{p}}D_{\rm{ds}}}
\end{equation}
and where $D_{\rm{dp}}$, $D_{\rm{s}}$, $D_{\rm{p}}$, and $D_{\rm{ds}}$ are the angular diameter distances from the deflector to the perturber, to the source, to the perturber, and from the deflector to the source. The flexion shift roughly measures the perturbations to the images of the source due to 3rd order terms from the perturber. Ultimately, what constitutes a "weak" or "strong" perturber is somewhat arbitrary, but there this is a clear trade-off between the improvement from including a particular perturber in the mass model and the amount of work required to do so. \cite{flexion_formalism} recommends using $\Delta_3 x = 10^{-4}\text{arcsec}$ as the cutoff between strong and weak perturbers to ensure a $<1\%$ bias on $H_0$.

\subsection{Comparison Datasets}

The first step in determining the value of $\kappa_{\rm{ext}}$ is comparing the field of interest to a large number of lines of sight from some large reference field. This comparison gives us an empirical estimate of the relative matter density of the lens field as compared to all lines of sight in the universe. The reference field should be large enough to avoid sampling bias. For a small comparison field (on the order of a few $deg^2$) statistical overdensities or underdensities may occur \citep[see for example][]{Fassnacht_overdense}. However modern survey datasets are available which cover hundreds to thousands of square degrees, allowing us to use a sufficiently large comparison field to avoid sampling bias (see section \ref{hsc_an} for a discussion of our choices for this analysis). Ideally the data for the reference field and lens field are taken by the same instrument and processed by the same analysis pipeline. This turns out to be the case for the analysis of J0924 discussed later in this paper, but will not be true in general. At a very minimum we seek data with at least one band in common, with deep enough observations to produce meaningful results. When comparing the survey dataset to our line of sight, we set a magnitude cut and remove any objects from both datasets fainter than this cut. Previous work \citep[see][]{Collett_2013} has suggested that $i < 24$ represents a good limit, as setting a fainter limit does not appear to meaningfully impact the results of this style of analysis. This limit is also bright enough to be well above the detection limits of modern sky surveys, ensuring reliable photometry.

Once the appropriate data are in hand, it is important to consider which objects should be used in the comparison. It is important to set a magnitude cut that will remove objects too close to the detection limit of the instrument for photometry to be reliable, while still leaving enough data to make robust estimates of the relative density. However setting too bright of a limit will result in having too few objects to compare to.

Additionally, we cut out all objects with a redshift greater than the redshift of the source quasar, as these objects will not affect the path of the light as it travels from the quasar to our telescopes.

\subsection{Weighted Number Counts of Lens Field}\label{wnc_theory}

Because the value of $\kappa_{\rm{ext}}$ cannot be directly measured, we first define tracer quantities that can be computed directly from the available data. By comparing the value of these quantities in the lens field to the value of the identical quantity computed for a large number of reference fields, we obtain an empirical estimate of the relative density of the line of sight of interest.  As a first approximation, we expect the greatest contribution to $\kappa_{\rm{ext}}$ from massive objects close to the center of the line of sight. The primary mass contribution in any line of sight will be dark matter halos, but these are not directly observable. To quantify the contribution from these weak perturbers, we compute weighted number counts of the visible structures (i.e. galaxies) in the lens field as compared to a large number of reference fields. This technique has been used extensively in previous work \citep{Fassnacht_2006_env,Suyu_2010, Greene_2013, strides_env} . To do this, we select a region of interest around the lens and compare it to a large number of identically-shaped fields selected at random from the reference survey. At each step, we compute the ratio of the weighted number counts for the galaxies in the lens field to the identical statistic computed in the given reference field. For the given step, the value of the weight is therefore:

\begin{equation}
W_i = \frac{\sum_{j}w_{j,\rm{lens}}}{\sum_{j}w_{j,i}}\end{equation}

Where $i$ indexes the reference fields, $j$ indexes the galaxies in a given field, and $w$ is the value of the weighted statistic for the given galaxy.

Following \citet{rusu_0435} we also consider a second style of weighting that improves our results. Instead of summing the value of weights for all objects in the field, we instead compute the weight for a given field as $w_{i,\rm{meds}} = n_i\overline{w_j}$ where $n_i$ is the number of galaxies in the given reference field and $\overline{w_j}$ is the median value of the weight for all galaxies in the reference field. Doing this helps avoid situations where single objects dominate the sum in a particular line of sight. This is especially important for weights involving stellar mass and the inverse seperation. In this scheme, the value of the weight for the given reference field is therefore

\begin{equation}
    W_{i,\rm{meds}} \equiv \frac{n_{\rm{lens}}\overline{w_{j}}_{, \rm{lens}}}{n_i \overline{w_{j,i}}}
\end{equation}

There has been discussion in the literature  about which are the best weights to consider for the most robust determination of $\kappa_{\rm{ext}}$ \citep{Greene_2013,rusu_0435, rusu_wfi}. In this work, we only consider a subset of the weights considered in previous works (see Table \ref{wnc_definitions}).

\begin{table}
\caption{Weighted Number Count Definitions}
\label{wnc_definitions}
\centering
\begin{tabular}{ccc}
\hline\hline
Name & Value & Symbol \\
\hline
Number Count & $w_j = 1$ & $w_n$ \\
Inverse Distance & $w_j = 1/r_j$ & $w_{1/r}$ \\
Potential & $w_j = m_j/r_j$ & $w_p$ \\
Redshift & $w_j = z_s \cdot z_j - z_j^2$ & $w_z$ \\
z/r & $w_j = w_{z,j}/r_j$ & $w_{z/r}$ \\
\hline
\end{tabular}
\end{table}

\subsection{Weighted Number Counts in Simulated Data}

In order to determine the posterior distribution of $\kappa_{\rm{ext}}$ for the given system, it is necessary to compare the weighted number counts to similar counts obtained from a reference field for which $\kappa_{\rm{ext}}$ is known. We use a simulated dataset for this purpose. The simulation must contain several components in order to be suitable for this analysis. First, it must contain catalogs of galaxies with known luminosity and redshift. Second, values of the external convergence must be measured at a suitably large number of points to be representative of the universe at large.  \citet{rusu_0435} examined possible biases from inferring $\kappa_{\rm{ext}}$ using the number counts method in the Millennium Simulation, and found these methods produced a good estimate of $\kappa_{\rm{ext}}$.  We discuss our choice of simulation further in section \ref{ms}. Because this technique involves ratios, much of the dependence on the simulation's underlying cosmological parameters should cancel out. However, ensuring this would require a second simulation with the attributes described. An exciting development in this space are the initial results from the MillenniumTNG \citep{millenniumtng}. Full weak-lensing convergence maps are planned, but not yet available.  

\subsection{From Weighted Number Counts to $\kappa_{\rm{ext}}$}\label{inference}

We now have weighted number counts for the lens field itself and for a large number of fields in a simulated dataset, each of which is associated with a value of $\kappa_{\rm{ext}}$. We seek to compute $p(\kappa_{\rm{ext}} | \textbf{d}$): the probability distribution of $\kappa_{\rm{ext}}$ given the data. We can replace this with the joint probability distribution of $\kappa_{\rm{ext}}$ \textit{and} the data as follows:

\begin{equation}
p(\kappa_{\rm{ext}} | \textbf{d}) = \frac{p(\kappa_{\rm{ext}}, \textbf{d})}{p(\textbf{d})} = \int dW_q \frac{p(\kappa_{\rm{ext}}, W_q, \textbf{d})}{p(\textbf{d})}
\end{equation}

After some work it can be shown \citep[see][]{rusu_0435}

\begin{equation}
    p(\kappa_{\rm{ext}} | \textbf{d}) = \int p_{\rm{sim}}(\kappa_{\rm{ext}} | W)p(W|\textbf{d}) \prod_{i}dW_i
\end{equation}

Where $p(W|\textbf{d})$ is the probability distribution of given weighting scheme given the data, and $p_{\rm{sim}}(\kappa_{\rm{ext}} | W)$ is the probability distribution of $\kappa_{\rm{ext}}$ in the simulated dataset, given a particular value of the weight. Here, it is assumed that the simulated dataset is the correct prior for the observable universe. In other words, that it correctly relates values of weights to values of $\kappa_{\rm{ext}}$

Additionally, a constraint on $\kappa_{\rm{ext}}$ based on $\gamma_{\rm{ext}}$ can be included, which accounts for the expected correlation between these two values. In general, $\gamma$ is a two-dimensional vector on the plane of the sky, but we use only the overall magnitude in our analysis. $\gamma_{\rm{ext}}$ is a parameter that can be fitted in the primary mass model of the lens. We use $\gamma_{\rm{ext}}$ just as we would a weighted number count distribution. For any range of values of $\gamma_{\rm{ext}}$, we can construct a histogram of the value of $\kappa_{\rm{ext}}$ for all lines of sight with values of $\gamma_{\rm{ext}}$ that fall in this range. We then weight the contribution from these lines of sight by $p(\gamma | \bold{d})$, which is the posterior on $\gamma$ inferred from the mass model.This provides a prior on $\kappa_{\rm{ext}}$ that can meaningfully affect the final result. In section \ref{r_d}, we present results both with and without using $\gamma_{\rm{ext}}$ as a constraint.

In previous work \citep[e.g.][]{rusu_0435}, this full probability distribution for each weight $p(W|\textbf{d})$ was replaced by a normal distribution centered on the median of the full weight distribution, with a width determined by examining measurement uncertainties. In practice, this width was much smaller than the width of the actual distribution. This is cheaper computationally, but ignores covariance between the various types of the weights. While the increase in computation time is significant, the majority of the important decisions in the analysis are made when we compute weighted number count ratios, and using this method does not meaningfully increase our time-to-result 

Since all the individual weights are measured at the same sequence of randomly-drawn fields, we can construct a full \textit{m}-dimensional probability distribution $p(W_s | \textbf{d}) = p(W_n, W_{1/r}, ...|\textbf{d})$ where \textit{m} is the number of weights being considered. We then explore this probability distribution when implementing the formalism described above. Formally, the posterior on $\kappa_{\rm{ext}}$ becomes:
\begin{equation}
    p(\kappa_{\rm{ext}} | \textbf{d}) = \int p_{\rm{sim}}(\kappa_{\rm{ext}} | W_s) p(W_s|\textbf{d})d^mW_sp
\end{equation}

When computing this quantity, we split the \textit{m}-dimensional probability distribution into $200^m$ \textit{m}-dimensional bins. We have also tested this procedure $100^m$ bins and see consistent results for the J0924 field. With significantly more bins, the computational time balloons and the number of fields in each bin drops significantly, even near the center of the distribution. The value of $p(W_s|\textbf{d})$ is simply the number of lines of sight in the reference survey that fall in this bin divided by the total number of lines of sight being considered. Given the large numbers of lines of sight we consider, the distributions are  smooth and it may be possible to model them explicitly and explore the distribution with an MCMC, but we save this for a future analysis. The exception to this is the pure unweighted number counts in a $45^{\prime\prime}$ aperture, due to the fact that the "weight" is integer valued and the number of galaxies in the aperture is relatively modest. 

For $p_{\rm{sim}}(\kappa_{\rm{ext}} | W_s)$ we construct a histogram of the measured values of $\kappa_{\rm{ext}}$ for all lines of sight from the simulated datasets that fall within the given bin, normalized by the number of lines of sight in the bin. Without this normalization, $\kappa_{\rm{ext}}$ values near the mean of the simulation will always be weighted more heavily because there are comparatively more of them. Our choice of simulation is discussed in section \ref{ms}

The inputs to our analysis code include the full weight distributions for the lens field, the weights computed from the simulated dataset, and the $\kappa_{\rm{ext}}$ maps for the simulated dataset for the source redshift. We iterate over the probability distribution discussed above, computing a histogram of the values of $\kappa_{\rm{ext}}$ in each \textit{m}-dimensional bin. The overall histogram is the sum of these histograms, weighted by the value of the weight distribution in that bin. 

\subsection{Comparison to Other Methods}

There are other methods for estimating $\kappa_{\rm{ext}}$ which have been explored as discussed below. Ultimately many of these techniques involve a trade-off between speed of analysis and precision of the final result. Our goal is explicitly to design and implement techniques that allow us to analyze hundreds or even thousands of lens systems in a reasonable amount of time, with the goal of combining full results (including lens modeling, and time delay measurements) from many lenses to make some final statement about the value of $H_0$. Because of this, a technique with slightly less constraining power for a single lens is tolerable if it can be performed and iterated on rapidly. A few other techniques for estimating $\kappa_{\rm{ext}}$ are discussed below. While all show promise, and are interesting for their own sake, none show a significant enough improvement to justify the increased time and complexity of analysis, at least in the context of our stated large-scale goals.

\subsubsection{Weak Lensing Analysis}

While the value of $\kappa$ at any given point on the sky is not observable, $\gamma$ \textit{can} be measured by looking a distortions in the shapes of galaxies in the line of sight. Assuming $\gamma$ can be measured, techniques such as those presented in \citet{Kaiser_1995} can be employed to reconstruct the underlying mass distribution. However this analysis requires extremely high quality data about the morphology of galaxies in the line of sight, as $\gamma$ is measured from the extremely small distortions that are present in the images of the galaxies. Obtaining such high quality data, as well as the overhead of analyzing it, represents a significant bottleneck. This analysis was performed in \citet{holicow_wl} on the same line of sight analyzed in \cite{rusu_0435}. Results between the two methods are consistent, with only modest improvement to precision for the weak lensing analysis.

\subsubsection{Explicit Modeling}

Building an explicit mass model of all perturbers (or at least, some larger sample) is attractive from a pure astrophysics perspective. This approach was explored in \citet{flexion_formalism} and demonstrated some success. However, the analysis assumed lines of sight with extensive spectroscopic coverage. While this may be true for some lines of sight, spectroscopy is expensive in time and resources and it is not obvious that the improvement is significant enough to justify this. Furthermore, explicitly modeling a line of sight requires making assumptions about the host halos of the galaxies, a potential source of additional bias.

\subsubsection{Machine Learning Methods}

An additional interesting approach using Bayesian graph neural networks (BGNNs)  is presented in \citet{gnn_kappa}. This approach was compared to a toy version of the analysis discussed here, using only a single summary statistic instead of a combination of several. The BGNN technique demonstrated greater precision and accuracy over using a single summary statistic on simulated data, however we estimate the difference would be much less significant if the comparison was done against the full line of sight analysis discussed in this work. That said, a more detailed comparison between these techniques would be welcome in the future. \citet{gnn_kappa} also demonstrated a meaningful bias for both techniques in more extreme fields ($\kappa < -0.05$ and $\kappa > 0.06$) due to a lack of similarly extreme fields in simulated datasets to compare to. This is worth examining further, but does not immediately suggest significant improvements from using the BGNN techniques, especially for extreme fields.

\section{Datasets, automation, and \texttt{lenskappa}}

As with many areas of astronomy, time-delay cosmography is grappling with datasets that are growing at unprecedented rates. There are dozens of known quad lenses which may be suitable for cosmographic analysis, but a full analysis has only been completed on a small fraction of them. Techniques for increasing the rate of analysis are therefore crucial to the continued success of the technique. 

While there are many open science questions in Time Delay Cosmography, much of the solution to the rate problem lies in software engineering rather than astronomy. When building for this kind of of analysis, we keep three key questions in mind:

\begin{itemize}
    \item What is the minimum amount of data required to complete a particular analysis step with sufficient precision?
    \item How much of the analysis can realistically be automated?
    \item Do we expect the analysis techniques to change significantly in the future?
\end{itemize}

The answer to these questions may not be independent. For example, a pipeline that uses less data at the expense of increasing uncertainties on individual systems may be tolerable if it significantly increases the rate at which these systems can be analyzed. The third question is also important. Writing flexible software packages that can easily be updated as analysis techniques evolve usually increase the time to first result, but substantially decrease \textit{average} time to result in the long run.

Among the various  steps required to fully analyze a lens system, the number counts technique discussed in this paper is likely the most straightforward to automate. It is mostly statistical, and the most difficult computational challenge is efficiently filtering a large dataset by location. Furthermore the survey datasets used in this analysis are quite robust, and many of the lens systems fall within one or more survey footprints. This makes the first question a non-issue, at least for a significant fraction of the lenses.

To that end, we introduce \texttt{lenskappa}\footnote{https://github.com/PatrickRWells/lenskappa} and \texttt{heinlein}\footnote{https://github.com/PatrickRWells/heinlein}. The goal of \texttt{lenskappa} is to build a tool capable of automating environment analysis to the greatest extent possible, while still providing sufficient flexibility to allow us to iterate on our current methods. \texttt{lenskappa} is in turn built on top of \texttt{heinlein}, which serves as a high-level interface to locally stored astronomical datasets. When computing weighted number counts in \texttt{lenskappa}, the core weighting loop consists of: 

\begin{enumerate}
    \item Select a region of interest from a large survey.
    \item Retrieve object catalog and auxiliary data for the region.
    \item Compute interesting quantities, using the data retrieved for the region.
\end{enumerate}

\begin{figure*}
\includegraphics[width=18cm]{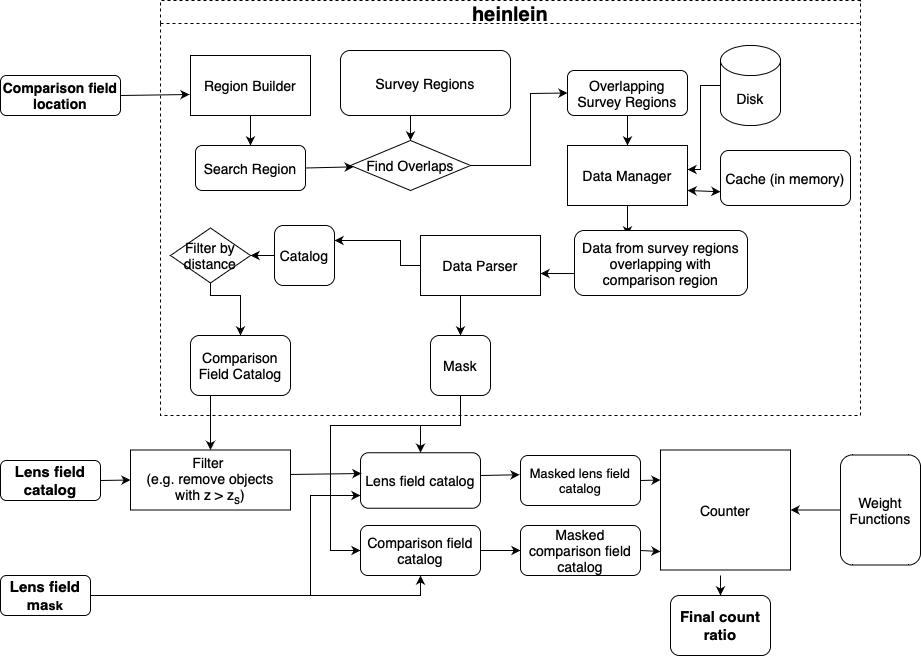}
\caption {Flowchart representing the process of computing a weighted count ratio given a location in a reference survey. Work done in \texttt{heinlein} is contained within the dotted box.}
\label{flowchart}
\end{figure*}

\noindent A single iteration of this loop is represented in Figure \ref{flowchart}

\texttt{heinlein} handles the middle step of this loop. It provides high-level routines for storage, retrieval, and filtering of large survey datasets, as well as intuitive interfaces for interaction between data types (for example, applying a bright star mask to a catalog). \texttt{heinlein} can perform a 120" cone search in the HSC dataset in around 1.5 seconds, without requiring the data be pre-loaded into memory. For later queries of nearby locations, the speed is improved by more than an order of magnitude through caching. This makes \texttt{heinlein} suitable both for interactive use and for the kind of analysis done in \texttt{lenskappa}. 

With data retrieval optimized, \texttt{lenskappa} focuses on allowing users to design and implement analyses that operate on large swathes of the sky. The techniques described in this paper, for example, could easily be adapted to build mass maps of the universe as seen in these surveys. However the goal is to work towards a tool which allows for much more flexibility, enabling \textit{any} analysis that involves calculations done on many small regions within a large astronomical survey. Other applications could include lens finding, though we note that image data is not yet supported in \texttt{heinlein.}

\texttt{lenskappa} includes several features to facilitate this including:

\begin{itemize}
  \item High-level API for defining analyses
  \item Plugin architecture for adding new capabilities without modifying the core code.
  \item Automatic support for any dataset supported by \texttt{heinlein}
\end{itemize}

\subsection{Analyzing modern astronomical survey datasets at scale}  One of the big challenges in doing analyses on these kinds of datasets is the need for the computing environments to be close to the data whenever possible. Querying over the internet is useful for assembling datasets, but is not a particularly good solution when analyzing a dataset at scale.  Many researchers will not have access to sufficient storage to store these datsets, and it is impractical to expect individual survey teams to provide computing resource for general use. The size of these datasets will enable next-generation analyses, but only with the development of next-generation tools running at scale, which will require computing infrastructure that may not be readily available to many researcher.

We support using cloud computing services to fill this gap. Commercial providers have expanded and matured by leaps and bounds over the last decade, and routinely handle storage and analysis tasks on datasets orders of magnitude larger than the ones being discussed here. Additionally, cloud computing technologies are significantly more accessible then on-premise technology: they require far lower startup costs and can be quickly scaled (to accommodate more users, or bigger jobs) without the bottlenecks that slow down the expansion of on-premise infrastructure. 

It is also possible to use cloud solutions as a supplement to already-existing on premise solutions. \citet{hepcloud} demonstrated this by analyzing data from the Compact Muon Solenoid experiment at massive scale.
 
In the future, we plan to develop \texttt{lenskappa} and \texttt{heinlein} tools that could be easily deployed onto services like these, providing quick and easy access to large survey datasets in addition to techniques for processing and analyzing that data. The datasets would be stored in the cloud, allowing users to deploy their analyses without worrying about the connection to the underlying dataset. Such an approach has been demonstrated by \citet{serverless_cms}, which enabled serverless access to $\texttt{ROOT}$ for quick analysis tasks on high energy datasets without the user having to manually retrieve the data. With some work, \texttt{heinlein} will enable this type of analysis by serving as a bridge between  computing infrastructure (which could be managed by individual researchers)  and the underlying data lake (which could be managed by the survey team). Abstracting away data retrieval will allow researchers to focus on what matters most: designing the analyses they wish to perform and interpreting results.

\section{Analysis of SDSS J0924+0219}

In this section, we discuss the analysis of the J0924 system, as performed by \texttt{lenskappa}.

\subsection{The Lens Field}\label{lens_field_sec}

\begin{figure*}
\includegraphics[width=16cm]{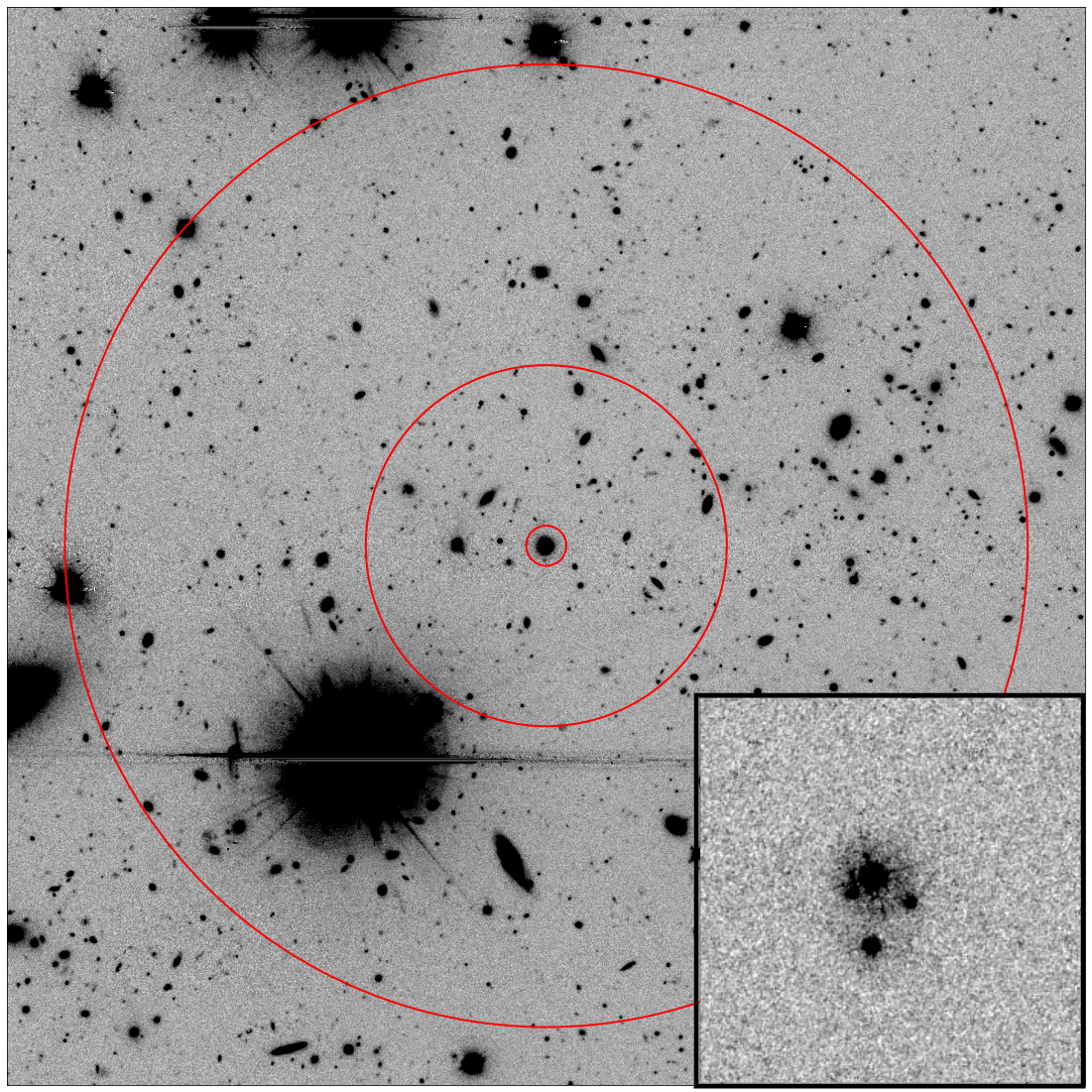}
\caption {Field around SDSS J0924, shown in HSC i-band. The red rings represent 5", 45" and 120" apertures respectively.  Bottom right: $10"$ cutout of the lens system from HST imaging.}
\label{j0924_image}
\end{figure*}

SDSS J0924+0219 is a quadruply lensed quasar first reported in \citet{j0924_discovery}. The quasar itself is at redshift $z = 1.523$, while the lensing galaxy is located at $z = 0.384$ \citep{lens_z}. Quadruply lensed quasars are particularly valuable for cosmographic analysis because it is in principle possible to measure 12 different time delays, though only three of these are independent. An image of the field can be seen in figure \ref{j0924_image}. This system was modeled in \cite{0924_model}

One particularly important feature of the lens field is the bright star located in the lower left. The star covers a reasonably large fraction of the overall area in the 120" aperture, undoubtedly covering several background objects and making photometry for objects very near it unreliable. We apply techniques that have previously been used in this kind of analysis to correct for this. This technique is discussed in more detail in section \ref{hsc_an}.

The lens field falls within the Subaru Hyper Surpime-Cam Strategic Survey Program footprint, and full color information is available for all objects in the relevant catalogs. We use these catalogs, including photometric redshifts, for all objects inside the field. Additionally we use the bright star masks and photometric redshift PDFs provided by the HSC team. 

Our analysis follows the same outline discussed in section \ref{overview}. We discuss the details particular to this lens field below. 

\subsection{The HSC Survey and Weighted Number Count Ratios}\label{hsc_an}

The Hyper Suprime-Cam Subaru Strategic Survey is a large survey program, aiming to cover roughly 1400 deg$^2$ of sky in five photometric bands (\textit{grizy}) down to  $i \sim$ 26, with deeper coverage expected in smaller regions of the sky \citep{hsc_design}. We base our analysis on the roughly 400 deg$^2$ that had coverage in all five bands as of the second data release \citep{hsc_dr2}. Data release 3 was made available while this paper was in preparation, but we do not consider it here.

The HSC Survey is a natural choice for a comparison dataset for this system because the J0924 field itself falls within the survey footprint. We therefore have both robust and comparable photometry. When computing weighted number count ratios, the basic approach is identical to the one outlined in the section \ref{wnc_theory}, with some specific adjustments:

We use objects with \texttt{r\_extendness\_value} $ = 1.0$, which selects galaxies. \citet{hsc_software} demonstrates that their algorithm for computing this quantity does a reasonably good job of selecting galaxies, though it may incorrectly classify some galaxies as stars, and vice-versa. However we note that the HSC-wide survey was intentionally selected to enable cosmological analyses by being in regions that are away from the galactic plane and low on dust extinction \citep{hsc_design}. This, combined with robust bright-star masks and our large sample size ensures unmasked stars do not significantly impact the final weighted number count ratios.

The full photometric redshift PDFs for all objects in the data release have been made available by the HSC team \citep{hsc_photoz}. They use two separate fitting algorithms, and results from both algorithms are included in the catalog. We compute weighted number count ratios using both sets of redshifts, and do not find a meaningful difference between the resultant distributions. As such, we use the "DEmP" redshift and stellar masses for our analysis \citep{dempa,dempb}.

When computing weighted number counts, we remove all objects closer than 5 arcsec from the center of the field following \cite{rusu_0435}. Objects this close to the center of the field are typically explicitly included in the mass model of the lens, and so we also remove them from the comparison fields to avoid biasing results.     

The HSC survey team makes available masks that represent areas of the sky where photometry may be unreliable or lacking due to the presence of bright stars \citep{hsc_bsm}. This is particularly important in our field due to the presence of the bright star that can be seen in Figure \ref{j0924_image} When iterating over the reference survey, we retrieve the bright star masks for each region being considered. We apply both these masks and the masks for the lens field itself to both catalogs at each weighting step. Doing this ensures that results are not biased if a given field in the reference survey has significantly more or less of its area covered by bright star masks. This procedure was first used in \citet{rusu_0435}

The 400 deg$^2$ of sky we consider here is separated into seven disconnected regions. Initially, we computed weighted number count ratios in five of these regions. This produces nearly-identical distributions, with the difference between the lowest and highest median being $0.1\sigma$ the standard deviation of the distribution, suggesting our fields are large enough to to avoid sampling bias. Based on this, we restrict our subsequent analysis to 135 deg$^2$ of sky located in the region $332^{\circ} < \text{RA} < 359^{\circ}$ and $-1.5^{\circ} < \text{Dec}< 6^{\circ}$. For each combination of aperture and limiting magnitude, we compute weighted count ratios at 100,000 randomly selected fields. 
\begin{figure}
\includegraphics[width=8cm]{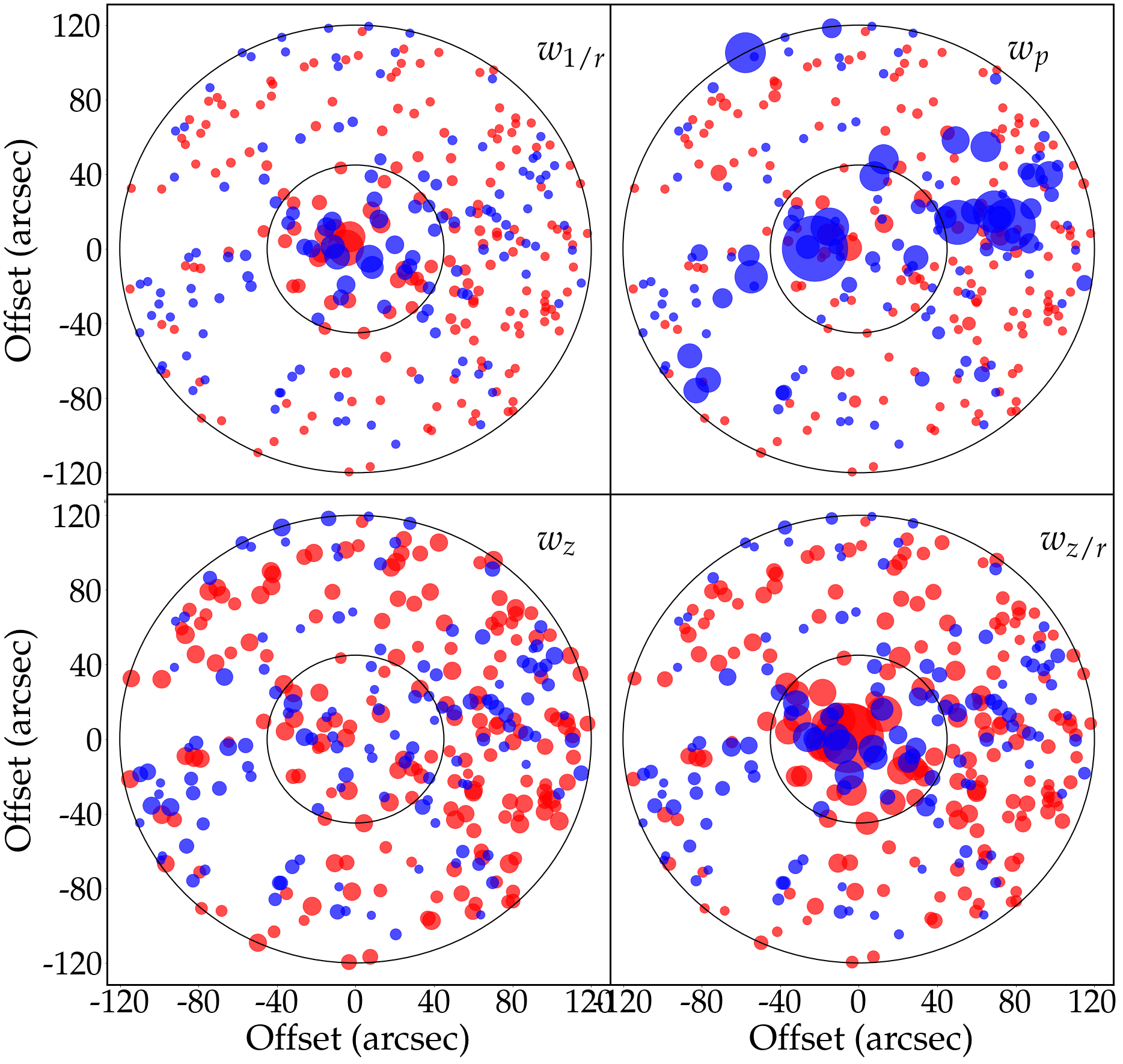}
\caption {Location and relative weights of all galaxies with $z < z_{\rm{s}}.$ The black circles represent the 45" and 120" apertures respectively. Blue objects are those with i < 23, while red objects are those with 23 < i < 24. Relative object weights of the object are represented by the size of the dot.}
\label{catalog}
\end{figure}

\subsection{Millennium Simulation}\label{ms}

The Millennium Simulation \citep{ms_main} is a dark matter only simulation split into 64 $4\times4\deg^2$ fields. After the original run was completed, synthetic galaxy catalogs were painted into the resultant halos by several teams. Following \citep{rusu_0435} we use the semi-analytic catalogues of \citet{sa_cats}. Additionally, \citet{Hilbert_raytrace} split each $4\times4\deg^2$ field into a grid of 4096x4096 points and used ray tracing to compute convergence and shear at each of these points in 63 redshift planes. These, combined with its large size, makes it an excellent choice for our analysis. In our analysis, we use redshift plane 36 with $z = 1.504$.

First, we compute the weighted number counts at a large number (order $10^6$) of equally spaced grid points in the Millennium simulation. For the $45"$ aperture, we place the fields $90"$ apart (snapped to the nearest grid point), while for the $120"$ aperture we place fields $60"$ apart. Both cases result in over 1.5 million lines of sight across the simulation, each of which has an associated value of $\kappa_{\rm{ext}}$ and $\gamma_{\rm{ext}}$ This differs from the same calculation for the lens field itself in a key way: the values reported are the total value of the weights at every point considered, rather than a ratio of values. To normalize, we divide weighted number count in each field by the median value for all lines of sight in our sample. The median value of the resultant distribution is therefore unity.

A key difference between the synthetic catalogs and the real data from the HSC survey is that in the Millennium Simulation catalog redshifts are exact. In previous work \citep[eg.][]{rusu_0435} this difference was accounted for by computing photometric redshifts for all objects in the Millennium Simulation, using the same pipeline that was used to compute the photometric redshifts in the comparison dataset. We are unable to use this technique in this case because the HSC survey photo-\textit{z} pipelines are not publicly available as of the preparation of this paper. Instead, we download the full catalog of training data used by the HSC. These are galaxies for which spectroscopic redshifts are available. We divide the objects in the test dataset into redshift and magnitude bins. For each galaxy in a bin, we compute the offset in the central value of the redshift $(z_{phot} - z_{spec})$, and take the median of these as our estimate of the redshift bias in that bin. Additionally we take the median value of $\sigma_z$ (as reported by the HSC photometric redshift pipeline) for all photometric redshifts in that bin. We take the the median value of $\sigma_z$ as our estimate of photometric redshift uncertainty. For each object in the Millennium Simulation catalog, we construct an artificial photemetric redshift PDF. The center of the distribution is the "actual" redshift of the object, offset by the the amount computed previously for the appropriate bin. The width of the distribution is the value of $\sigma_z$ computed for the same bin. 

At each weighing step in the Millennium Simulation, we then sample from these "photometric redshift" PDFs when computing weighted number counts. For each line of sight, we sample from each of these PDFs 50 times. This produces 50 separate catalogs (all with slightly different values for object "redshifts"), and we compute weighted number counts for each one. We find that this process produces no meaningful change to our final inference for $\kappa_{\rm{ext}}$, even when a weight that depends on the redshift is taken into account. However this process massively increases the amount of data output by our code, and therefore significantly increases the amount of time required to do the final inference on $\kappa_{\rm{ext}}$. This does suggest that photometric redshift uncertainties do not have a significant impact on the inferred value of $\kappa_{\rm{ext}}$, but we plan to explore this more completely in the future.

\section{Results and Discussion}\label{r_d}
Our weighted number counts for the lens field include a total of five weights ($w_n$, $w_{1/r}$, $w_p$, $w_z$, $w_{z/r}$, see Table \ref{wnc_definitions}), two apertures (45" and 120"), two limiting magnitudes (23 and 24), and two summing techniques (pure sum and medians). For brevity, we report only the medians of these distributions in Table \ref{wnc_medians}. The full distributions, along with the analysis code used to produce the posterior distributions for $\kappa_{\rm{ext}}$ are available on github. \footnote{https://github.com/PatrickRWells/J0924-analysis}. A visual catalog of the objects in the field and their relative weights can be seen in Figure \ref{catalog}. 

\begin{figure*}
\includegraphics[width=16cm]{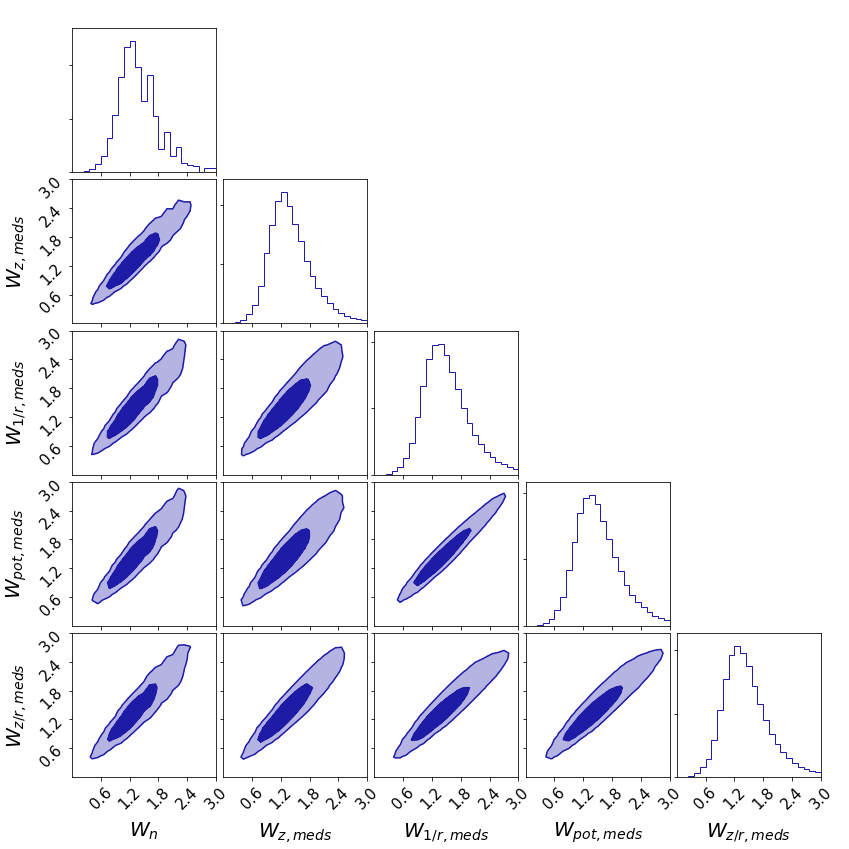}
\caption {2D histograms of for each possible pair of the weight number count ratios considered in this work, computed with a $45"$ aperture and limiting magnitude $i < 24$. The inner and outer contours represent 68 and 95\% confidence intervals, respectively.}
\label{2d_distro}
\end{figure*}

The weighted number count ratios suggest the SDSS J0924 field is mildly overdense as compared to the universe as a whole. This overdensity is significantly more obvious when considering the $45$" aperture. This is quite reasonable; as the size of the aperture increases, the density of field will approach the density of the universe as a whole. See Figure \ref{2d_distro} for an overview of the weights considered in this work.

It is however less obvious why the value of the weights seem to depend on the limiting magnitude, with the median values for $i < 23$ being significantly higher than for $i < 24$. This may suggest that the quality of the lens field catalog is poor below magnitude 23,  We remind the reader that the $5$" region around the lens itself is masked when computing weighted number counts.
\begin{table}
\caption{Medians of Weighted Number Counts for J0924}
\label{wnc_medians}
\begin{tabular}{ |c|c|c|c|c| } 
\hline\hline
{Weight} &  i<23, 120"&  i<24, 120" &  i<23, 45" &  i<24, 45" \\
\hline
$w_n$&    1.04 &    1.04 &   1.42 &   1.35 \\
$w_{1/r}$       &    1.17 &    1.09 &   1.57 &   1.35 \\
$w_p$      &    1.19 &    1.10 &   1.56 &   1.36 \\
$w_z$        &    1.03 &    0.99 &   1.56 &   1.37 \\
$w_{z/r}$         &    1.21 &    1.07 &   1.74 &   1.37 \\
$w_{1/r,meds}$  &    1.08 &    1.01 &   1.59 &   1.45 \\
$w_{p,meds}$ &    1.12 &    1.01 &   1.50 &   1.47 \\
$w_{z,meds}$   &    1.03 &    1.00 &   1.50 &   1.36 \\
$w_{z/r, meds}$
&    1.10 &    1.04 &   1.63 &   1.40 \\
\hline

\end{tabular}
\end{table}

Based on the above, we would expect our final value of $\kappa_{\rm{ext}}$ to be somewhat positive. However the strength of the external shear of the field, reported in \citep{0924_model} is $0.017^{+0.001}_{-0.003}$. This places it significantly below the median (and, in fact, mean) for all lines of sight in the Millennium Simulation. For each combination of limiting magnitude and aperture we compute $\kappa_{\rm{ext}}$ for a number of different combinations of weights:

\begin{itemize}
    \item Pure (unweighted) number counts, combined with each of the remaining weights individually.
    \item Pure number counts and inverse distance weights, combined with each of the remaining weights individually.
    \item For each of the above cases, we run the kappa inference both with and without the constraint from $\gamma$.
\end{itemize}

The number count ratios used for this analysis are listed in Table \ref{wnc_medians}. We do not mix distributions obtained using our two different weighting techniques. 

We also repeat this for each while using the median weighting scheme rather than the sum scheme. All together, this leaves us with 112 individual histograms for the value of kappa. 

We find that the choice of specific weights is less important than the number of weights being considered. In all cases, inferring $\kappa_{\rm{ext}}$ with three weights instead of two tightens the resultant distribution, but does not significantly affect the central value. \cite{rusu_0435}  found best results when combining $w_n$ and $w_r$ with one additional weight and the constraint from $\gamma_{\rm{ext}}$. We find the same here. Specifically, the central value of the distribution does not change meaningfully based on the choice of the third weight, but we find slightly tighter constraints when using $w_p$.

We also find that results are consistent between apertures, but find that a brighter value of the limiting magnitude results in a noisier posterior on $\kappa_{\rm{ext}}$. Because so many objects fall between magnitude 23 and 24, removing these objects results in significantly noisier weighted number counts which translate to the final inference on $\kappa.$
\begin{figure}
\includegraphics[width=9cm]{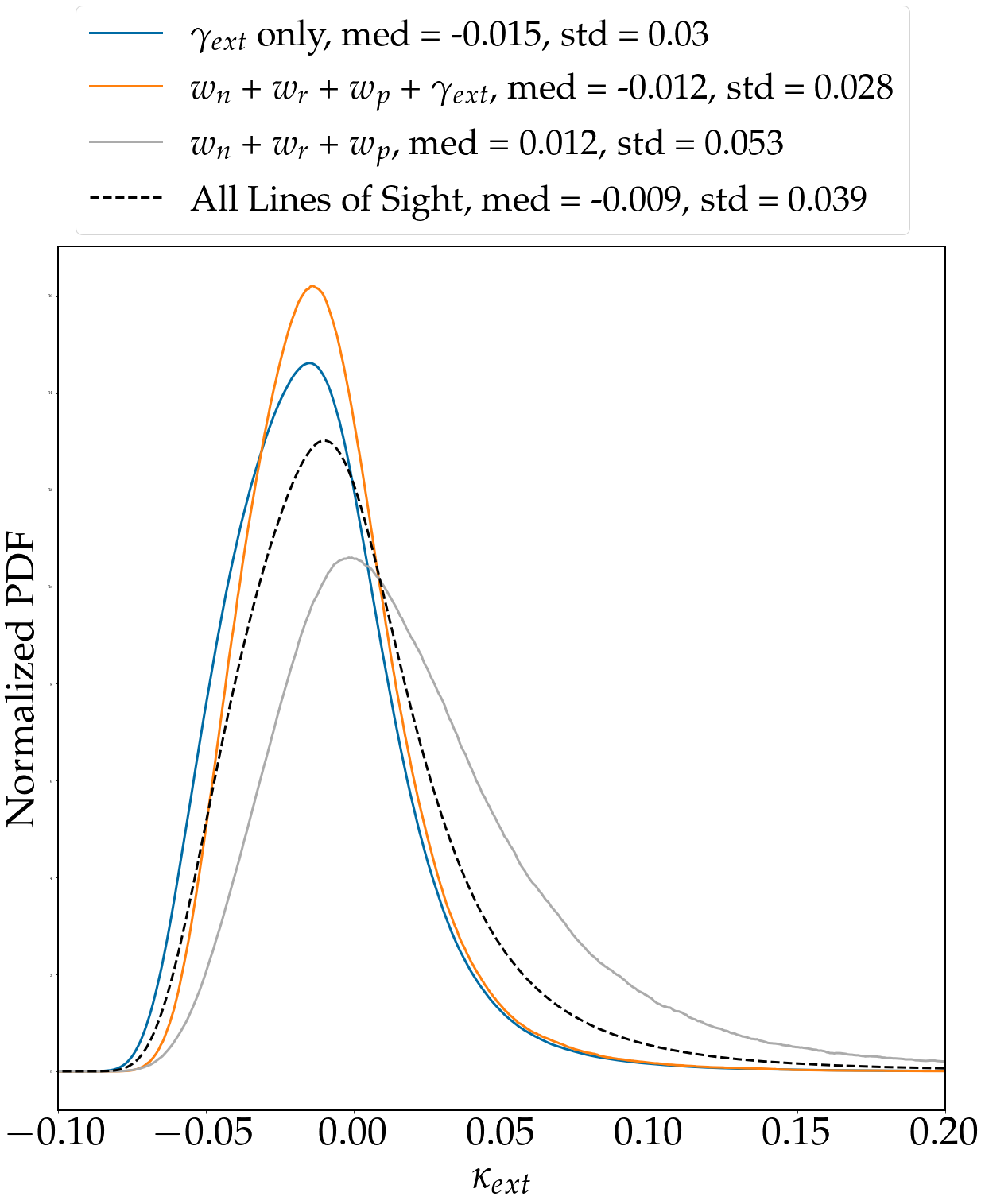}
\caption{Comparison of smoothed posterior on $\kappa_{\rm{ext}}$ for $w_n + w_r + w_p$ without $\gamma_{\rm{ext}}$, $w_n + w_r + w_p$ with $\gamma_{\rm{ext}}$, and $\gamma_{\rm{ext}}$ alone.}
\label{kappa_compare}
\end{figure}
Considering only $\gamma$ as a constraint leads to a median value on $\kappa_{\rm{ext}}$ of -0.015. As a result, including $\gamma_{\rm{ext}}$ as a constraint significantly lowers the central value of our distributions, though it also tightens the distribution. However we note this shift is fairly modest as compared to the width of the distribution. This is similar to the result seen in \citet{rusu_0435}, though in that case the inferred value of the shear was significantly closer to the median in the Milennium Simulation of 0.028, and the shift of the central value was not as significant. We find our tightest constraints on $\kappa_{\rm{ext}}$ through a combination of $w_n$, $w_{1/r}$ and $w_{p}$ combined with constraints from $\gamma_{\rm{ext}}$. This leads us to a final value of $\kappa_{\rm{ext}}$ of -0.012 with a width $\sigma_{\kappa} = 0.028.$ Without the constraint from $\gamma_{\rm{ext}}$, we obtain a median value of 0.012 with $\sigma_{\kappa} = 0.053.$ Full posteriors for these combinations can be found in Figure \ref{kappa_compare}. We use a $45"$ aperture and limiting magnitude of $i < 24$ for these results.

\section{Conclusions and Future Work}

In this paper, we have discussed the current state of the line of sight number counts technique for environment analyses in time delay cosmography. We have introduced two main improvements to previous iterations of the analysis. First, our packages \texttt{lenskappa} and \texttt{heinlein} make designing and running these analyses much quicker than before, in addition to making it much simpler to add additional survey datasets. This will accelerate the pace of future analyses, and enable population-level analyses of lens environments. Additionally, we have made use of the entire distributions of weighted number counts, which accounts for covariance between weights and is generally more robust than just exploring a small region around the medians. We have applied these techniques to the J0924 field, and found that this field is a fairly typical line of sight, with a slightly negative median value of $\kappa_{\rm{ext}}.$  

\subsection{Future Development of \texttt{lenskappa}}

Our primary goal for this project has been to build a software tool that can quickly and reliably analyze weak perturbers along lines of sight to strong gravitational lenses. We have accomplished this goal, but we plan to extend the capabilities of \texttt{lenskappa} to include tools for analyzing strong perturbers and coherent structures (such as galaxy groups). These objects must be handled individually, and require very different analysis tools. However we see a significant advantages to being able to do full environment analyses within a single software package.

\subsection{Future Analyses}

Leveraging the capabilities of \texttt{lenskappa}, we hope to better understand lines of sight to gravitational lenses on a population level. \citet{Fassnacht_overdense} and \citet{Wong_overdense} have shown that lenses seem to fall in preferentially overdense lines of sight, but that this overdensity seems to be confined to the immediate surroundings of the lens itself. \texttt{Lenskappa} gives us the tools to perform these population-level analyses quickly, with the freedom to adjust and re-run the analysis as needed. As a first step, we hope to complete an analysis with 4-5 new lens systems, as well as 2-3 systems that have previously been analyzed to check our results.

As an additional check, we would like to analyze a large number of non-lens lines of sight. This would ensure there are no biases introduced by comparing distributions based on real galaxy catalogs to those obtained from the synthetic catalogs in the Millennium Simulation.

Recently, \citet{gnn_kappa} demonstrated the use of Bayesian Graph Neural Network to estimate the value of $\kappa_{\rm{ext}}$ in a simulated dataset. Their method out-performs a simplified version of the analysis performed here that uses only a single weight. Further work may demonstrate the ability of the technique to match or even outperform the weighted number counts technique, but a more complete comparision will need to be performed to assess this.

\begin{acknowledgements} 
P.W. and C.D.F. acknowledge support for this work from the National Science Foundation under Grant No. AST-1907396.
\linebreak
P.W. thanks Liz Buckley-Geer, Anowar Shajib, and Simon Birrer for useful discussions throughout the preparation of this manuscript.
\end{acknowledgements}
%
%
\bibliographystyle{aa}
\bibliography{main}

\end{document}